\documentclass[sigplan,10pt]{acmart}
\renewcommand\footnotetextcopyrightpermission[1]{}
\AtBeginDocument{%
  }

\usepackage{xspace}
\usepackage{subcaption}
\usepackage{cleveref}
\usepackage{multirow}
\usepackage{arydshln}
\usepackage{enumitem}
\usepackage{tikz}
\usepackage{xcolor}
\usepackage{soul}
\usepackage{graphicx}
\usepackage{amsmath}
\usepackage{booktabs}
\usepackage{microtype}
\usepackage{circledsteps}
\newcommand*\encircle[1]{\Circled{#1}}
\usepackage{algorithm}
\usepackage[noend]{algpseudocode}
  {\begin{itemize}
    [leftmargin=0cm,
     itemindent=.3cm,
     labelwidth=\itemindent,
     labelsep=0pt,
     parsep=3pt,
     topsep=2pt,
     itemsep=1pt,
     align=left]
  }%
  {\end{itemize}}

\definecolor{dyryu_blue}{rgb}{0.3, 0.7, 1}

\newcommand{\dyryu}[1]{} 

\newcommand{\ours}{\textsc{CacheSmith}\xspace}
\newcommand{\warmup}{\textsf{WarmUp}\xspace}
\newcommand{\probe}{\textsf{Probe}\xspace}
\newcommand{\patch}{\textsf{Patch}\xspace}

\usepackage[utf8]{inputenc}
\usepackage[T1]{fontenc}
\usepackage{kotex}

\setcopyright{rightsretained}
\acmDOI{}
\acmISBN{}
\acmConference{}{}{}
\acmSubmissionID{} 

\renewcommand\footnotetextcopyrightpermission[1]{}
\renewcommand{\shortauthors}{Anonymous}

\graphicspath{{figures/}}

\newcommand{\includefigure}[2][]{%
    \IfFileExists{#2}{%
        \includegraphics[#1]{#2}%
    }{%
        \fbox{\parbox[c][0.25\textheight][c]{0.95\linewidth}{\centering\textcolor{red}{[MISSING FIGURE]}}}%
    }%
}

\newcommand{\PP}[1]{\vspace{2px}\noindent{\bf #1.}}

\def\Snospace~{\S{}}

\newcommand{\orangebox}{%
  \raisebox{0.05ex}{%
    {\setlength{\fboxsep}{0pt}\setlength{\fboxrule}{0.4pt}%
     \fbox{\textcolor{orange}{\rule{1ex}{1ex}}}}%
  }%
}

\newenvironment{myitemize2}%
  {\begin{itemize}
    [leftmargin=.3cm,
        itemindent=0pt,
        labelwidth=.15cm,
        labelsep=.15cm,
        parsep=0pt,
        topsep=2pt,
        itemsep=1pt,
        align=left]
  }%
  {\end{itemize}}

  {\begin{enumerate}
	[leftmargin=0cm,itemindent=.5cm,labelwidth=\itemindent,
		labelsep=0pt,
		parsep=1pt,
		topsep=1pt,
		itemsep=3pt,
		align=left]
  }%
  {\end{enumerate}}

\usepackage{dsfont}
\newcounter{takeaway}

\newcommand{\takeaway}[1]{\refstepcounter{takeaway}\PP{Takeaway~\thetakeaway}\label{#1}}

\renewcommand{\footnoterule}{\kern-3pt\hrule width 0.4\columnwidth\kern 2.6pt}

\begin{document}
\pagestyle{plain}

\title{\ours: Accurate RAG Serving under Latency SLOs with Adaptive KV Reuse}
\author{Xinyue Ma}

\affiliation{
  \institution{POSTECH}
  \country{}
}

\author{Damin Lee}
\affiliation{
  \institution{POSTECH}
  \country{}
}

\author{Aashaka Shah}
\authornote{Currently at Mirendil.}
\affiliation{
  \institution{Microsoft Research}
  \country{}
}

\author{Myeongjae Jeon}
   \authornote{Corresponding author: \href{mailto:mj.jeon@postech.ac.kr}{mj.jeon@postech.ac.kr}}
\affiliation{
  \institution{POSTECH}
  \country{}
}
\begin{abstract}

This paper presents \ours, a KV cache reuse system for RAG prefill that
maximizes quality-gated SLO attainment. While context caching lowers prefill
latency by reusing precomputed KV states for retrieved chunks, it misses
cross-chunk attention that a full prefill would compute over concatenated
inputs. Prior selective recomputation schemes tackle this problem but enforce
static, one-size-fits-all configurations that fail to effectively handle
request heterogeneity or dynamic latency budgets. \ours decomposes selective
recomputation into a three-phase workflow of \warmup, \probe, and \patch that
can be dynamically configured to exploit a wide range of cost-accuracy
trade-offs. At serving time, an SLO-aware scheduler in \ours forms batches to
mitigate inter-request interference during layer-by-layer prefill, while a
budget-aware planner configures recomputation phases across the batch to
maximize the number of requests satisfying both generation quality and latency
targets. Both components make fast, solver-free decisions over profiled cost-accuracy bands. 
On four benchmarks across varying arrival rates and burstiness levels, \ours
improves quality-gated SLO attainment by up to 27.5\%, over four state-of-the-art baselines.
\end{abstract}

\maketitle

\section{Introduction}
\label{sec:intro}

Production LLM serving platforms now handle billions of daily queries across
domains~\cite{chatgpt_scale}. Many of these deployments rely on
retrieval-augmented generation (RAG)~\cite{rag}
to provide a knowledge base as context to the LLM to reduce hallucination~\cite{hallucination_survey}.
Specifically, at serving time, the RAG system ranks candidate \emph{text
chunks} (e.g., document passages or web pages) retrieved from an external
corpus by their relevance to the query~\cite{bm25, dpr, colbert, contriever}.
It then prepends the top-ranked chunks to the input prompt, so that the model
conditions its response on explicit query-specific evidence.

However, this comes
at a substantial latency cost. The prepended chunks
routinely
extend prompt lengths to tens of thousands of tokens per query~\cite{longrag}, all
of which must flow through prefill.
Because prefill cost scales quadratically with input length, it dominates the
time-to-first-token (TTFT) under RAG. \emph{Context caching} reduces this
cost by prefilling each chunk in advance and reusing the resulting key-value
(KV) cache across queries that retrieve the same chunk~\cite{RAG-DCache,
cache-craft, cachegen, lmcache}.
Under context caching, the KV cache of each chunk is precomputed in isolation, and thus captures attention only within itself, while completely missing the cross-chunk attention over the full context.
Without cross-chunk attention, the model struggles on queries that
require synthesizing facts or contextual dependencies across multiple
chunks~\cite{cacheblend, epic, kvlink}.


\emph{Selective recomputation} has emerged as a promising direction to
resolve this latency-accuracy tension~\cite{cacheblend, epic, fusionrag,
prophetkv}. The idea is to recompute KV states for a small fraction of
critical tokens over the concatenated chunks, in order to restore missing
cross-chunk attention while reusing the rest of the cache. This builds on the
empirical observation that most precomputed KVs already closely match
full-prefill ground truth, with marked divergence concentrated on only a
sparse set of tokens~\cite{cacheblend}. Existing work evaluates individual
context tokens for selective recomputation via two distinct selection
signals: (1) \emph{KV deviation}, which measures how far each token's cached
KV state drifts from full prefill~\cite{cacheblend, epic}, and (2)
\emph{query-context interaction}, which ranks tokens by their attention
strength to the query~\cite{fusionrag, prophetkv}. Both lines aim to recover
full-prefill quality without paying the quadratic latency cost of a fresh
prefill. However, our analysis shows that they do not adapt well to diverse
serving setups for two reasons.


First, prior methods adopt a one-size-fits-all scheme with a fixed
configuration for the selection signal and recomputation plan. For example,
CacheBlend~\cite{cacheblend} uses a predetermined set of early layers to
identify top-deviated tokens for
subsequent layers, whereas ProphetKV~\cite{prophetkv} enforces a rigid
two-pass mechanism to evaluate query-context interaction across all layers
before recomputing tokens in the second pass. In both cases, token selection
is strictly governed by a predefined ratio for every request. However, user
requests differ inherently in how much cross-chunk attention they must
recover to reach full-prefill quality. This difference can be
\emph{structural} (e.g., chunk count and chunk sizes) as well as
\emph{contextual}:
a multi-hop query combining facts scattered across chunks requires heavy
recomputation, while a simple lookup query finding its answer within a single
chunk needs minimal intervention. Consequently, existing designs often
under-recompute critical tokens and degrade generation quality,
or incur latency overheads and violate stringent TTFT
SLOs (\autoref{fig:char-isocost}(a)).
In practice, failing to meet either requirement renders the request
unsuccessful~\cite{mfs, proteus,tabi}, directly degrading what we term \emph{quality-gated SLO
attainment}.


Second, selective recomputation schemes today lack coordinated budget
allocation across concurrent requests. Requests, once batched, proceed
through the model layer by layer in lockstep to amortize weight-loading
overheads~\cite{orca, vLLM, sglang, sarathi-serve}. Thus, one request's execution cost
directly propagates into the latency of all other requests in the batch. This
makes batch-level scheduling particularly challenging, as pending requests
are highly heterogeneous: beyond varying in structural and contextual
complexity, they enter execution with disparate latency slacks after queuing.
In~\autoref{sec:char-batch}, we show that current schedulers routinely co-batch mutually
interfering requests, and thus inevitably push urgent requests past their
SLOs or starve recomputation-heavy requests during lockstep execution.

To this end, we present \textbf{\ours}, a
serving system for context-cached RAG designed to optimize
quality-gated SLO attainment by satisfying both TTFT SLOs and generation
quality targets.
We make the following contributions:

$\bullet$ \textbf{Execution abstraction.} To overcome one-size-fits-all
practices, we decompose monolithic selective recomputation into a modular,
three-phase operational workflow: \warmup that forms robust selection signals
during initial execution, \probe that reads them to rank context tokens, and
\patch that selectively recomputes top-ranked tokens across target layers.
This abstraction unifies prior techniques into a concrete design space
governed by four tunable knobs (selection signal, warm-up depth, probe length,
and patch coverage) and systematically exposes their cost-accuracy
trade-offs. Critically, it unlocks previously unexplored plans, such as
single-pass query-context interaction that suffices for contextually simple
queries, laying the foundation for request-level adaptation in \ours.


$\bullet$ \textbf{Quality-gated SLO-aware scheduler.} Building on this
abstraction, \ours optimizes quality-gated SLO attainment across requests.
The challenge is that this objective tightly couples three interdependent
decisions: which requests to batch, what budget the batch runs under, and
which configuration each request executes. \ours decouples this complexity
across (1) a \emph{budget-aware recomputation planner} that maps any granted
budget to the configuration maximizing accuracy and (2) an \emph{SLO-aware
scheduler} that forms batches and allocates budgets by evaluating the
remaining latency slacks of all queued requests rather than dispatching them
in na\"ive FIFO order. These two layers interface via each request's
\emph{budget band}, the profiled budget range where recomputation yields high
accuracy gains. Because scheduling over these pre-profiled bands requires
only simple arithmetic at runtime, \ours imposes minimal overhead on the
serving critical path.


$\bullet$ \textbf{Phase-aware KV pipelining.} 
\ours integrates operational
phase-aware prefetching that effectively overlaps host-to-GPU KV transfers with prefill
compute.
It first exploits \warmup layers, which run
full prefill without reading cached context, as an ample compute window to
mask initial loading latency for \probe. 
In addition, when evaluating
candidate plans, \ours accounts for potential transfer stalls under deeper
\probe layers, where sharper selection signals reduce required patch coverage
but increase KV loading volume from greater cache reuse. If such stalls
threaten SLO violations, \ours steers toward alternative plans that
stay within the latency budget without sacrificing accuracy.

We implement \ours on top of vLLM~\cite{vLLM} and evaluate it on two NVIDIA RTX PRO 6000 GPUs, each with 96GB of memory. We serve LLaMA-3.1-8B on a single GPU and Qwen-3-32B model on 2 GPUs with tensor parallelism. We evaluate four tasks: PassageRetrieval and MuSiQue from LongBench~\cite{longbench}, and VariableTracking and Multi-valueRetrieval generated following RULER~\cite{ruler} under
varying arrival patterns, SLO constraints, and data transfer bandwidth. By co-optimizing accuracy and the SLO, \ours improves quality-gated SLO attainment by up to 27.5\% and cuts P99 TTFT by up to 86\% over four
state-of-the-art methods: EPIC~\cite{epic}, ProphetKV~\cite{prophetkv}, CacheBlend~\cite{cacheblend}, and FusionRAG~\cite{fusionrag}.
\section{Background}
\label{sec:background}



LLM inference proceeds in two stages. \emph{Prefill} runs a forward pass over
the input prompt to build its per-token KV cache and produce the first output
token. \emph{Decode} then generates subsequent tokens one at a time, with
each attending to the KV cache of all preceding tokens. Under RAG, the
prefill stage processes multiple external text chunks (e.g., document
passages, web articles, or code snippets) in addition to the user query.
While essential for generation quality, these retrieved chunks drastically
inflate input length and prefill cost, making it challenging to satisfy
time-to-first-token (TTFT) SLOs. In this section, we provide background and
discuss existing efforts to address these challenges.

\subsection{RAG Context Caching}
\label{sec:bg-caching}


In RAG workloads, 
retrievals exhibit strong skewness toward a small set of
popular chunks~\cite{hedrarag, quake, vectorliterag}.
Therefore, na\"ively prefilling these shared chunks from scratch for every
request incurs significant redundant computation. To avoid this,
\emph{context caching} precomputes chunk KV caches in isolation and stores
them in host memory or storage~\cite{ragcache, RAG-DCache}. When a request hits a
cached chunk, its KV cache is transferred to GPU memory instead of being
fully recomputed. 

\begin{figure}
    \centering
    \includegraphics[width=\linewidth]{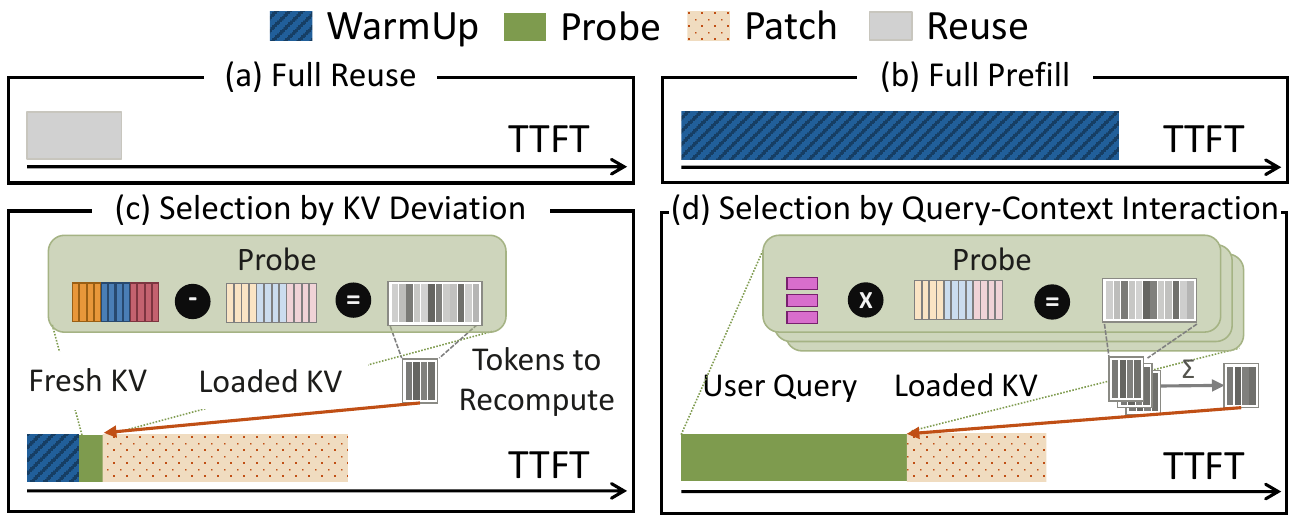}
    \vspace{-2em}
    \caption{Full Reuse, Full Prefill, and one for each family.}   
    \label{fig:methods}
    \vspace{-1.2em}
\end{figure}

However, context caching introduces a crucial fidelity gap. Precomputed on a
per-chunk basis, the cached KV contains no attention to preceding tokens in
the prompt. For instance, if a sequence of chunks $[A, B, C]$ is prepended to
a user query, chunk $C$'s KV lacks all attention weights directed toward $A$
and $B$. This missing \emph{cross-chunk} context is request-specific, as
chunk $C$ requires different cross-chunk attention in request $[D, E, C]$
than in $[A, B, C]$. Without the ability to precompute dynamic chunk
combinations offline, two straightforward options exist, as shown
in~\autoref{fig:methods}: (a) \emph{Full Reuse} simply
stitches together the original per-chunk KV caches and leaves the missing
context unrecovered at the risk of degrading response quality, and (b)
\emph{Full Prefill} computes the exact ground-truth KV over the entire prompt
at the latency cost that caching was meant to avoid.

\subsection{Selective Recomputation}
\label{sec:bg-methods}


\emph{Selective recomputation} strikes a middle ground between full reuse and
full prefill by updating a small fraction of context tokens over the
concatenated input to recover missing cross-chunk attention
online~\cite{cacheblend, epic, fusionrag, prophetkv}.
Prior work falls into two categories based on how it defines token
importance (\autoref{fig:methods}): (c) \emph{KV deviation} (KV-dev), which measures the divergence of a
token's cached KV from its full-prefill state, and (d) \emph{query-context
interaction} (QCI), which quantifies the attention the user query directs to
each context token. Methods within each family share this core definition but
differ in their runtime estimation strategies.


\PP{Selection by KV deviation} CacheBlend~\cite{cacheblend} serves as a prime
example of direct KV deviation estimation. It takes a single forward pass
over $L$ layers through three steps: (1) executing a full prefill on the
first layer to establish the ground-truth KV states, (2) measuring per-token
deviations from the cached KV across all $N$ context tokens (incurring an
$O(N)$ selection cost), and (3) recomputing only the top-$k$ tokens with the
largest deviations across the remaining $L-1$ layers. In contrast,
EPIC~\cite{epic} approximates deviation statically by token position,
observing that initial tokens in a chunk absorb disproportionately large
attention due to the attention sink phenomenon~\cite{streamingllm, attention_sink}.
EPIC thus simply recomputes a fixed number of initial tokens per chunk,
avoiding the $O(N)$ runtime selection overhead. However, it can miss critical
high-deviation tokens located deeper within chunks that CacheBlend captures.

\PP{Selection by query-context interaction} While KV deviation focuses on
intrinsic KV drift, QCI evaluates how relevant context tokens are to the
incoming query. ProphetKV~\cite{prophetkv} directly measures this interaction
via a pilot forward pass: $n$ user query tokens attend to $N$ cached context
tokens across all $L$ layers to score each context token by query attention,
incurring an $O(nNL)$ cost. It then fuses these per-layer scores and
recomputes only the top-$k$ tokens in a second full pass.
FusionRAG~\cite{fusionrag} reduces this cost by approximating the
interaction. Its pilot pass runs a query-only forward pass across all $L$
layers, computing self-attention over $n$ tokens at $O(n^2L)$ without the
cached context KV. Only at the final layer does the query attend to the
cached context to score QCI at $O(nN)$, in place of ProphetKV's all-layer
average.
The total selection cost is thus reduced to $O(n^2L+nN)$, which is dominated
by $O(nN)$ when $N>nL$, cutting ProphetKV's $O(nNL)$ cost by a factor of $L$.
Note that evaluating QCI solely at the final layer overlooks query-context
interactions in many preceding layers, which can degrade selection
accuracy~\cite{qcfuse}.

Although rooted in similar principles, existing approaches adopt static,
isolated designs. None examine how they perform under dynamic serving
workloads and whether they are suitable for the operational goals of
production systems.



\section{Static Reuse Under Dynamic Workloads}
\label{sec:char}

A serving workload is dynamic in two ways: incoming requests are
heterogeneous, and load
fluctuates over time. Neither can be predicted precisely. Under such a
workload, prior static selective recomputation methods fail on both counts.
Across requests, the recomputation required to restore the missing
cross-chunk attention varies widely
(\autoref{sec:char-need}), and the
optimal selection signal differs across query types
(\autoref{sec:char-signal}).
Moreover, a static configuration cannot adapt to each request's remaining
SLO slack, causing more SLO violations as load surges
(\autoref{sec:char-batch}). This section systematically establishes these
limitations
to motivate our workload-adaptive design.

\subsection{Request-Dependent Recomputation Demand}
\label{sec:char-need}



Recomputation demand is driven by two distinct factors: structural (chunk
boundaries where cross-chunk attention is lost) and contextual (reasoning
required by the query)~\cite{prophetkv}.
To quantify how both factors shape recomputation demand, we evaluate
PassageRetrieval~\cite{longbench} under CacheBlend by varying context length
from 10 to 40 random passages across a fixed set of 100
queries\footnote{6,000 passages prefilled in isolation, drawn from
PassageRetrieval~\cite{longbench}.}. For each request, we sweep the patch
coverage of CacheBlend (the fraction of context tokens recomputed) from full
reuse (0\%) to full prefill (100\%) in steps of 5\% and log the \emph{minimum
sufficient coverage} required to generate the correct answer. This setup
decouples the structural impact of chunk count from contextual variation
across individual queries.

\begin{figure}[t]
    \centering
    \includegraphics[width=\linewidth]{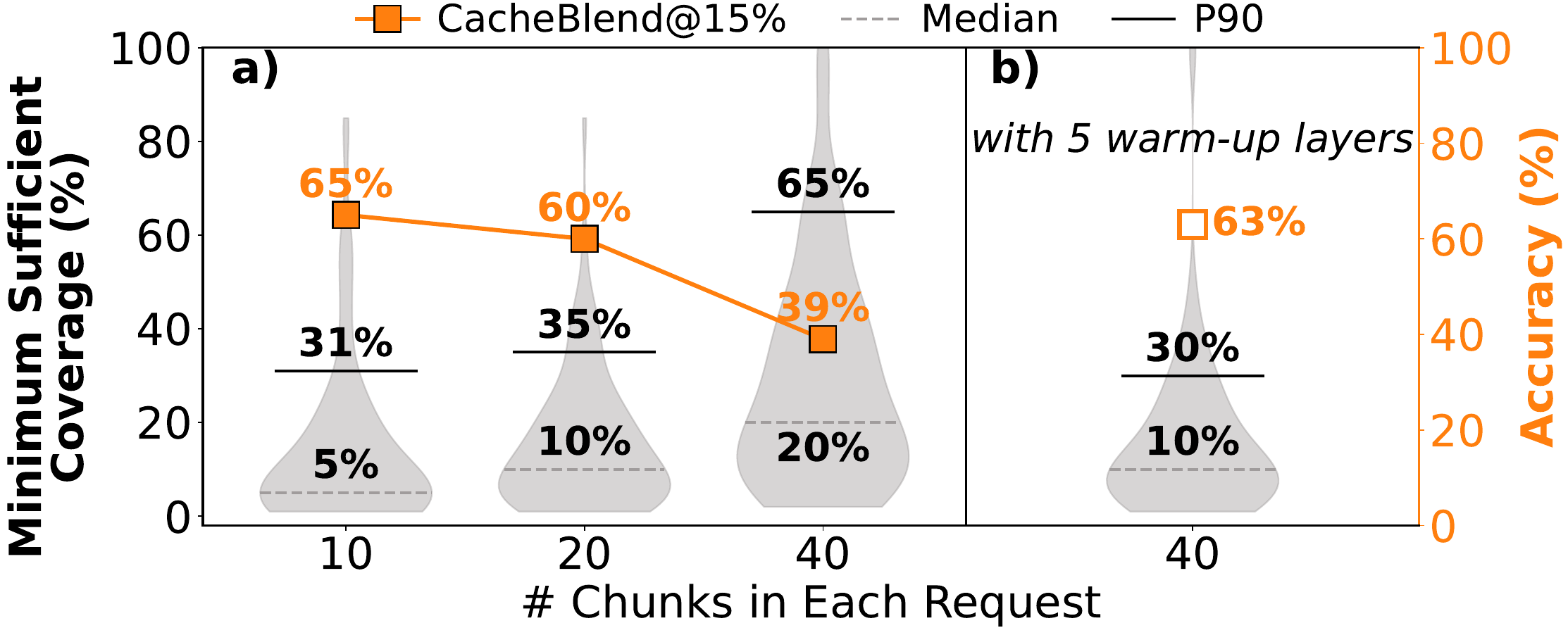}
\vspace{-2em}
\caption{Minimum sufficient coverage across passage counts in
PassageRetrieval. Violins show coverage distributions while the orange line
tracks CacheBlend@15\% accuracy. (b) adds five
warm-up layers to the 40-passage group.}
    \label{fig:char-minimum}
\vspace{-1.2em}
\end{figure}

As \autoref{fig:char-minimum}(a) shows, structural scaling sharply degrades
the accuracy of CacheBlend@15\% (\orangebox) from 65\% at 10 passages to 39\%
at 40 passages. To
satisfy 90\% of requests (the p90 line), 35\% coverage suffices for 20
passages, but longer requests with 40 passages demand 65\%. Moreover,
contextual differences introduce substantial intra-group variance.
Even with the same passage count, requests in each group exhibit minimum
sufficient coverage ranging from near zero to over 80\% while forming a
pronounced long tail. These results highlight the need to tailor
recomputation coverage individually rather than relying on a static ratio.


\PP{Improving ranking via warm-up} To keep overhead minimal, CacheBlend
inspects KV deviation using only the first layer. However, this choice yields
a noisy ranking that inevitably prioritizes irrelevant tokens and
necessitates higher coverage to reach the correct answer. We observe that
deeper layers capture richer cross-chunk interactions~\cite{pyramidkv,pyramidinfer}
and produce sharper rankings. As shown in \autoref{fig:char-minimum}(b),
computing the first five layers as a full-prefill warm-up on the 40-passage
group drops the median minimum sufficient coverage from 20\% to 10\% and the
p90 from 65\% to 30\%. This boosts CacheBlend@15\% accuracy from 39\% to
63\%.


\takeaway{tk:ranking} Across requests, structural and contextual heterogeneity
creates wide variations in required coverage. Within each request, warm-up
depth and patch coverage form a direct compute trade-off: each warm-up layer
executes full prefill over all tokens, consuming compute that could otherwise
expand patch coverage in later layers. In practice, a request can achieve
target accuracy through various combinations of warm-up depth and coverage,
defining a unique minimum-cost operating point. Prior fixed configurations
leave this optimization space unexplored.

\subsection{Task-Dependent Selection Signal}
\label{sec:char-rank}
\label{sec:char-signal}

\begin{figure}[t]
    \centering
    \includegraphics[width=\linewidth]{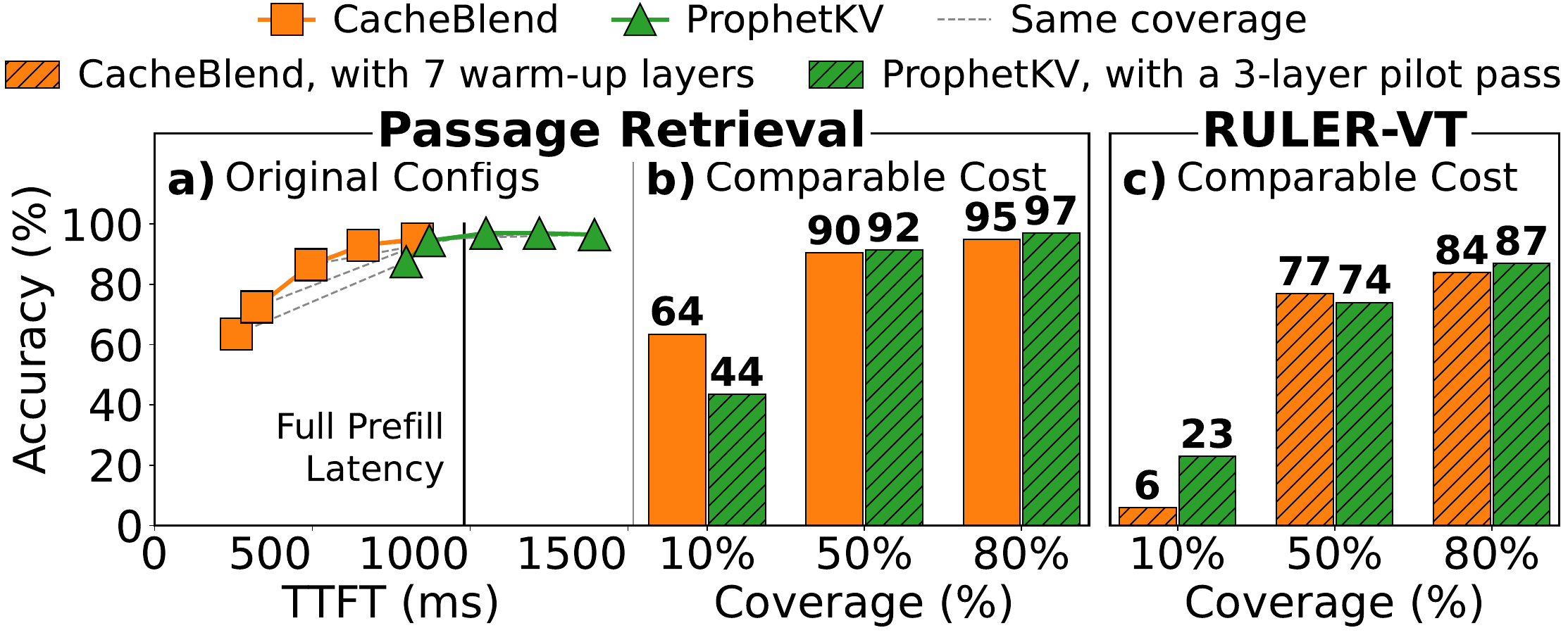}
\vspace{-1.5em}
\caption{Selection signals compared on 200 requests from PassageRetrieval. (a)~Original configurations with different coverages. (b,~c)~Accuracy under configurations with comparable cost on PassageRetrieval and RULER-VT.}
    \label{fig:char-signal}
\vspace{-1em}
\end{figure}

Existing work pays fundamentally different costs to collect token selection
signals. For instance, CacheBlend inspects intermediate KV states only after
first-layer execution, while ProphetKV runs a dedicated pilot pass to probe
and accumulate QCI signals across all layers. \autoref{fig:char-signal}(a)
compares ProphetKV with CacheBlend on patch coverages from 10\% to 80\% using
LLaMA-3.1-8B~\cite{llama3} on a single NVIDIA RTX PRO 6000 GPU. ProphetKV achieves
higher accuracy at all coverages, but its all-layer probe
incurs substantial latency overhead, causing TTFT to exceed even full-prefill
latency beyond 40\% coverage.


\PP{Signal efficacy under iso-latency} ProphetKV's accuracy gain is partly an
artifact of
its heavy probing overhead. To isolate signal efficacy from probing costs, we
re-evaluate both signals under an equivalent TTFT budget for each coverage by
adjusting CacheBlend's warm-up depth (1$\rightarrow$7) and ProphetKV's probed
layers (32$\rightarrow$3). As shown in \autoref{fig:char-signal}(b), at 80\%
coverage, both methods achieve comparably high accuracy
(97\% vs. 95\%), because recomputing
most tokens satisfies the demand of almost all requests. At 10\% coverage,
however, CacheBlend delivers 20\% higher accuracy than ProphetKV. This gap
stems from attention sparsity: QCI excels when query-context interaction
concentrates on a small set of tokens, whereas PassageRetrieval requires
spreading attention across the entire context to evaluate candidate passages
against the query and identify the matching one.
In contrast, RULER-VT~\cite{ruler} exhibits high attention concentration:
queries trace variable assignments through alias chains, so only tokens along
the reference chain matter. On this workload, ProphetKV's QCI probing
captures these key tokens effectively, yielding 17\% higher accuracy than
CacheBlend at the same 10\% coverage (\autoref{fig:char-signal}(c)).

\PP{Takeaway 2} No single token selection signal dominates across workloads, as
interaction patterns range from globally dispersed to highly localized.
Furthermore, because signal probing itself adds to prefill latency, choosing
a signal without accounting for its overhead can exhaust the TTFT budget and
starve downstream patch coverage.

\begin{figure*}[t]
    \centering
    \includegraphics[width=\linewidth]{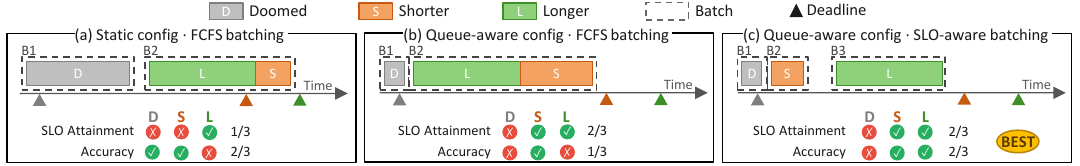}
 \vspace{-1em}
\caption{An illustrative example where a doomed (D), a shorter (S) and a longer (L) request are served under three
policies, pairing a static or queue-aware configuration with FCFS or SLO-aware
batching. }
    \label{fig:char-batch}
 \vspace{-1em}
\end{figure*}

\subsection{Load-Dependent Latency Budget}
\label{sec:char-batch}


Under fluctuating load, queueing delay can consume part of a request's TTFT
budget. Since an SLO violation degrades user experience regardless of
generation quality~\cite{sarathi-serve, distserve}, recomputation must adapt
to the request's remaining SLO slack. Crucially, a batch's budget is shared
due to lockstep model execution, and every second the batch spends delays the
requests queued behind it.


\autoref{fig:char-batch} illustrates the need to coordinate batching and
budgeting with a minimal example of three requests: a doomed request with
slack too small even for full reuse, an urgent short request, and a long
request requiring more recomputation. A static configuration (a) computes for
the doomed request anyway, exhausting the slack of subsequent requests. A
queue-aware configuration under FCFS batching (b) commits the doomed request
to full reuse to drain the queue sooner, but co-batches requests with
mismatched demands under a shared budget, underserving the long request while
overspending on the short one. In contrast, serving them separately (c)
tailors budgets to individual demands and deadlines, satisfying two of the
three requests. Throughout the paper, we count a request as served only when
it meets both its TTFT SLO and quality target, defining the fraction of such
requests as \textit{quality-gated SLO attainment}.

\PP{Takeaway 3} Under dynamic load, co-batching requests with mismatched
demands can compromise either TTFT SLOs or generation quality. Maximizing quality-gated SLO attainment therefore requires
coordinating batching with dynamic recomputation budgeting.

\begin{figure}[t]
    \centering
    \includegraphics[width=\linewidth]{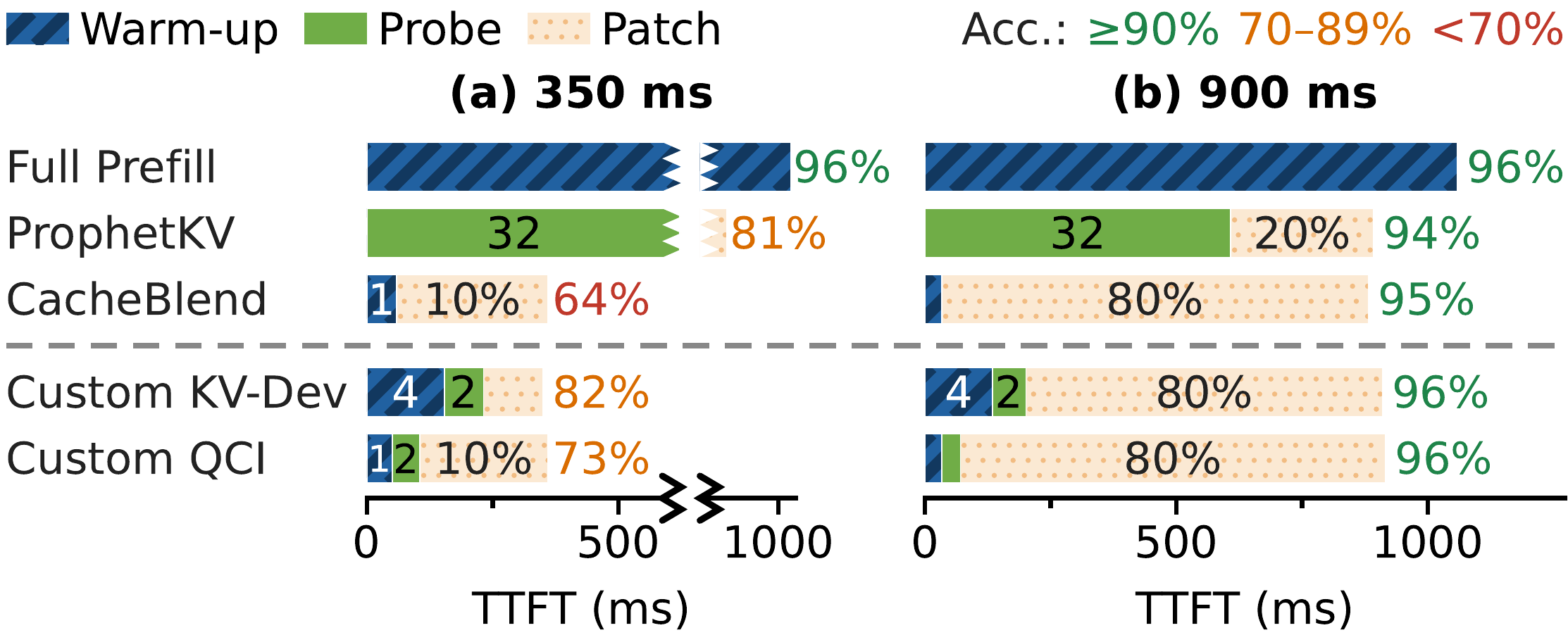}
\vspace{-1.5em}
\caption{Configs with varying selection signal, warm-up and probe layers, under budgets of 350\,ms (a) and 900\,ms
    (b).}
    \label{fig:char-isocost}
\vspace{-1em}
\end{figure}

\section{Workload-Adaptive Recomputation}
\label{sec:char-space}

Our characterization in~\autoref{sec:char} lays out three requirements for an
effective selective recomputation scheme:
\begin{myitemize2}
  \item It must dynamically alternate between KV deviation and QCI
  according to workload interaction patterns.
  \item It must expose intermediate execution phases as knobs:
  warm-up depth defers signal extraction to deeper, more reliable layers,
  while probe length bounds QCI overhead.
  \item It must bound the fraction of tokens to recompute to fit within the
  remaining TTFT budget.
\end{myitemize2}

Furthermore, warm-up and probing naturally complement each other. Warm-up
computes exact KV states in initial layers to recover ground truth at chunk
boundaries. Probing then accumulates query-context relevance over these
sharpened representations, which KV deviation alone cannot capture. We
illustrate the benefits of combining both knobs under TTFT budgets with a
running example below.


\PP{Execution abstraction} To formalize the recomputation design space, we
define an execution abstraction parameterized by a four-tuple $(S, w, p, r)$. A configuration executes in three phases: \warmup computes $w$ layers in
full, \probe accumulates the selection signal $S$ over $p$ further layers,
and \patch recomputes the top-ranked tokens up to a coverage $r$. The first
two phases determine token ranking, while coverage defines the fraction of tokens to be patched. 
Under this abstraction, prior methods are merely tied to static
points: $(\text{KV-dev}, 1, 0, 15\%)$ for CacheBlend and $(\text{QCI}, 0, 32,
20\%)$ for ProphetKV on a 32-layer model.


\PP{Example} \autoref{fig:char-isocost} compares full prefill, ProphetKV, and
CacheBlend against two custom plans (Custom KV-dev and Custom
QCI) under tight and ample TTFT budgets on PassageRetrieval. These custom
plans exercise both warm-up and probe for each signal, with optimal parameter
values found via grid search. Under the tight budget
(\autoref{fig:char-isocost}(a)), CacheBlend reaches only 64\% accuracy, while
ProphetKV misses the deadline at any coverage due to its probing cost.
In contrast, Custom KV-dev $(\text{KV-dev}, 4, 2, 5\%)$ achieves 82\%
accuracy and matches ProphetKV at half its TTFT. Custom QCI instead trades
accuracy for latency to achieve meaningful quality-gated SLO attainment where
ProphetKV fails. Under the ample budget
(\autoref{fig:char-isocost}(b)), all recomputation schemes converge,
indicating that our parameterized space provides effective operating points
across different TTFT budgets.


\PP{Design objectives} To realize this execution abstraction in an
end-to-end serving system, we target three objectives:
\begin{myitemize2}
  \item \textbf{Dynamic co-tuning.} Select the optimal
  configuration $(S, w, p, r)$ per batch according to request demands to
  maximize quality-gated SLO attainment (§3.1--§3.2).
  \item \textbf{Queue-aware budgeting.} Form a batch and set
  its budget to pursue accuracy gains while accounting for the queuing delays
  imposed on waiting requests (§3.3).
  \item \textbf{Lightweight online decisions.} Keep decision overhead negligible,
  as budgeting and knob co-tuning lie on the critical path and count against
  the TTFT budget.
\end{myitemize2}

\section{\ours Design}
\label{sec:design}

\begin{figure}
    \centering
    \includegraphics[width=0.95\linewidth]{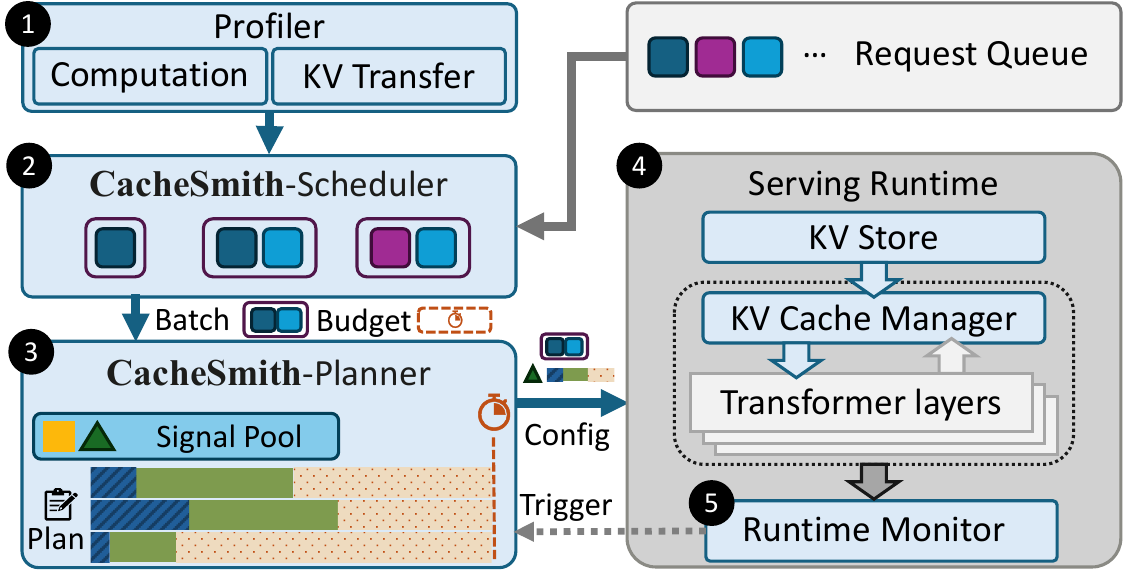}
    \caption{The overall architecture of \ours. 
    }
    \label{fig:arch}
    \vspace{-1.5em}
\end{figure}

We now present \ours, a budget-aware KV cache reuse system that optimizes
quality-gated SLO attainment (QSA) for RAG serving. A request counts toward
QSA only when it meets both its TTFT SLO and its quality gate:
\begin{equation}
  \mathrm{QSA} \;=\; \tfrac{1}{N}\sum\nolimits_{i=1}^{N}
  \mathds{1}\!\left[\,\mathrm{TTFT}_i \le \mathrm{SLO}_i
  \;\wedge\; q_i \ge \tau_q\,\right],
  \label{eq:qsa}
\end{equation}
where $q_i$ is the output quality of request $i$ under its task-specific
criterion and $\tau_q$ is the target quality gate.
Although forming a batch, setting its budget, and configuring recomputation
are interdependent, searching this joint space online is prohibitively
expensive on the serving critical path. \ours separates these decisions
between \ours-Scheduler and \ours-Planner (\autoref{fig:arch}).


The \textbf{profiler} \encircle{1} runs offline to measure computation and KV
loading costs for the deployed model under the server's compute capacity and
storage bandwidth. At runtime, upon completing each batch,
\textbf{\ours-Scheduler} \encircle{2} (\autoref{sec:design-sched}) forms the
next batch and sets its budget based on the remaining slack of queued
requests. To make informed decisions, \ours-Scheduler consults \ours-Planner
to obtain each candidate batch's \emph{budget band} (the latency range where
accuracy can be recovered without wasting compute resources) and ranks
candidate batches by how many requests can be placed inside their bands. Once
the batch and budget are decided, \textbf{\ours-Planner} \encircle{3}
(\autoref{sec:design-plan}) promptly navigates the configuration space under
the given budget to select $(S, w, p, r)$ that pushes the most requests past
the quality gate (\autoref{sec:char-signal}). The \textbf{serving runtime}
\encircle{4} (\autoref{sec:design-kv}) executes the chosen configuration
layer by layer, while its \textbf{KV cache manager} asynchronously prefetches
upcoming KV states to overlap data transfer with ongoing computation. During
execution,
if latency drift occurs to the plan and risks violating the budget,
\textbf{runtime monitor} immediately re-invokes \ours-Planner to adjust the
plan for the remaining layers.

\subsection{Recomputation Planner}
\label{sec:design-plan}
 

Given a batch and its assigned latency budget by \ours-Scheduler
(\autoref{sec:design-sched}), \ours-Planner selects an execution
configuration that maximizes aggregate request quality. Because requests in a
batch execute synchronously layer by layer, they must share the structural
ranking knobs: selection signal $S$, warm-up depth $w$, and probe length $p$.
In contrast, coverage is configured per request $r_i$, as each \patch layer
recomputes a different set of tokens for each request using its own cached
context and query tokens. \ours-Planner thus needs to strike the right
balance between refining ranking quality (via $S, w, p$) and expanding
per-request recomputation coverage (via $r_i$) within the assigned budget.


\PP{Choosing $S$ at runtime} As explained in~\autoref{sec:char-signal}, the
better selection signal depends on whether context attention is localized
(favoring QCI) or dispersed (favoring KV-dev).
To detect this online, \ours runs the first layer as an initial \warmup ($w
\ge 1$), and \ours-Planner measures the concentration of each request's QCI
scores via normalized entropy $H$ over the context tokens. Requests are
routed to KV-dev if $H > \theta$ (dispersed attention), and to QCI otherwise.
Because requests in a batch execute synchronously in lockstep, the batch
adopts the majority signal among its requests. Once the signal $S$ is
settled, \ours-Planner determines the remaining configuration knobs under the
cost-accuracy characteristics of the chosen signal.

\begin{figure}
    \centering
    \includegraphics[width=\linewidth]{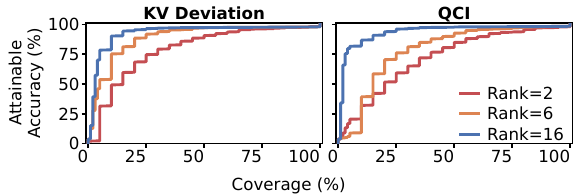}
\caption{The fraction of requests whose
minimum sufficient coverage is met for rank depths 2, 6, and 16.}
\label{fig:profile}
\vspace{-1em}
\end{figure}


\PP{Ranking-coverage trade-off} \ours-Planner defines the total layers
dedicated to ranking (\warmup and \probe) as the \emph{rank depth}. Deeper
ranking refines token selection to reduce required coverage
(\autoref{fig:char-minimum}(b)), but consumes budget needed for patching.
Moreover, per-request demand varies with a heavy tail
(\autoref{fig:char-minimum}(a)) and cannot be known upon arrival. \ours
therefore profiles the demand distribution offline and uses it as a proxy to
guide runtime decisions.

Because \ours-Planner needs to compare candidate rank depths relatively, it
does not require precise per-request demand prediction: the ability to rank
the candidates suffices. For a given coverage, the distribution yields the
fraction of requests whose demand is satisfied, which \ours-Planner
interprets as a request's probability of passing the quality gate. This makes
direct comparison across candidate rank depths possible: for instance, four
ranking layers might pass 70\% of requests, while five layers afford only
68\% despite sharper ranking. \ours-Planner thus iterates over candidate rank
depths and selects the one maximizing expected passing requests.

\ours profiles demand distributions by rank depth rather than by $(w,p)$
pairs, as the internal split between warm-up depth and probe length
empirically has negligible impact on accuracy. To build this profile
efficiently, we fix $p{=}1$, measure only three anchor depths (5\%, 20\%, and
50\% of model layers) using 500 synthetic requests from PassageRetrieval, and
interpolate between them (\autoref{fig:profile}). As shown in
\autoref{sec:eval}, this lightweight synthetic profile generalizes well even
to unseen tasks. In production, however, serving systems can periodically
refine the distribution using runtime execution traces to more faithfully
reflect real-world workload shifts.

\PP{Transfer-aware cost model} \ours-Planner determines the split between
\warmup ($w$) and \probe ($p$) online, as the execution cost depends heavily
on where the KV resides across storage tiers (e.g., host memory, NVMe SSD, or
remote stores). The layer types exercise storage differently: \warmup runs
pure computation without loading KV states, \probe requires its target KV
resident before execution, and \patch loads only reused tokens. While \probe
can reach a given depth with less computation than \warmup, placing it early
shifts KV transfer demand forward before sufficient \warmup compute has
executed to hide transfer latency. Therefore, when KV states reside on slow
tiers, initial \warmup layers are essential to provide an ample compute
window that masks full KV transfers for subsequent probes.

%

To find the optimal split, \ours-Planner estimates batch TTFT by walking
layers sequentially while accounting for compute-transfer overlap: layers
execute in order when their KV states are ready. The cost model takes two
inputs: layer computation times measured offline and effective KV transfer
bandwidth tracked by the runtime monitor. For each candidate $(w,p)$ pair,
the planner thus identifies the maximum feasible coverage fitting the budget
and picks the pair yielding the highest coverage. Crucially, when KV transfer
lags behind computation, the GPU idles until the next layer's KV arrives.
\ours exploits these I/O stalls to recompute lightweight early tokens (e.g.,
attention sinks~\cite{streamingllm}), affording higher coverage than a
compute-only model would project.

It is worth noting that transfer bandwidth can drop below its profiled value
during batch execution (e.g., due to I/O contention) and risk a budget
violation. In this case, the runtime monitor triggers \ours-Planner to
re-evaluate feasible coverage across the remaining layers.

\PP{Coverage allocation across requests} Since coverage is relative to
context length, an allocated token yields higher marginal coverage for
shorter requests. Under a fixed batch token recomputation budget,
\ours-Planner optimizes per-request coverage to maximize the expected number
of requests meeting the quality gate. Starting from a uniform coverage, the
planner iteratively shifts tokens from the request with the smallest marginal
loss in passing probability to the one with the largest marginal gain. It
terminates when no further transfer improves the total expectation.

\PP{Budget band} The trade-off above does not hold when the budget is too low
or too high. With a high budget, accuracy converges toward full prefill under
every configuration (\autoref{fig:char-isocost}b),
meaning a simple static configuration performs comparably to any optimized
split. With a low budget, ranking deeper than one \warmup and one \probe
layer takes compute away from an already scarce coverage.
In this case, accuracy converges toward full reuse, and the compute spent on
ranking is wasted. \ours-Planner therefore sets the budget \emph{floor} at
the TTFT of $(S, 1, 1, r_{\min})$ and the \emph{saturation point} at that of
$(S, 1, 1, r_{\max})$; the two define the batch's budget band. By default, we
set $r_{\min} = 5\%$ and $r_{\max} = 50\%$, and we provide a sensitivity
analysis on the budget range in~\autoref{sec:eval:ablation}. \ours-Planner delivers this
band to \ours-Scheduler as a contract: the scheduler invokes the planner only
when a batch budget falls within this band; otherwise, it bypasses the
planner and directly applies a static default configuration.


\subsection{SLO-Aware Scheduler}
\label{sec:design-sched}


A late response is unacceptable regardless of quality, just as a low-quality
response is unacceptable regardless of latency. Thus, \ours allocates compute
to neither requests missing their SLOs nor those failing the quality gate. To
optimize both dimensions, \ours-Planner enforces the gate, while
\ours-Scheduler maximizes SLO attainment across requests. Specifically, at
each batch boundary, \ours-Scheduler aims to maximize the number of requests
meeting their SLOs over the entire queue rather than the next batch alone. In
doing so, it treats the budget band returned by the planner as an accuracy
guarantee: a request placed inside its band has its quality secured as far as
its budget allows. As summarized in \autoref{alg:scheduler}, the scheduler
takes request deadlines, GPU KV cache limits, and candidate budget bands as
input to produce the next batch, its recomputation budget, and its KV cache
layout.

\PP{Budgeting for attainment} \ours weighs two key considerations. First,
increasing a batch's budget improves its quality but delays waiting requests.
Second, while lock-step execution naturally constrains a batch's budget by
its most urgent request, \ours can selectively permit an individual SLO
violation when the compute time reclaimed from that request allows more
requests to meet their budget bands.


To reconcile these trade-offs, \ours-Scheduler plans one batch at a time.
Planning the entire queue at once is computationally prohibitive and easily
invalidated by dynamic request arrivals. Instead, \ours-Scheduler evaluates
candidate batches against a conservative model of the remaining queue: after
the candidate batch runs, all waiting requests are assumed to execute
first-come-first-served (FCFS) at their budget floors. This pessimistic view
avoids expensive multi-batch combinatorial search while preventing
over-allocation to the current batch. Under this projection, \ours-Scheduler
iterates through candidate batch compositions. For each candidate, it raises
the budget until a request would miss its deadline, exceeding the saturation
point only while every waiting request still clears its budget floor. A
candidate is then scored by the total requests meeting their budget bands,
both in the batch and across the modeled queue. \ours-Scheduler stops at the
first composition that satisfies all requests, or otherwise selects the one
that satisfies the most.

\begin{algorithm}[t]
\algrenewcommand\algorithmicdo{}
\algrenewcommand\algorithmicthen{}
\caption{\ours-Scheduler}
\label{alg:scheduler}
\begin{algorithmic}[1]
\State \textbf{Input:} queue $Q$; KV capacity $M$
\State \textbf{Output:} batch $C$, budget $b$, KV window $\ell$
\State $\mathcal{B} \gets \textsc{FormBuckets}(Q)$
\For{$C \gets \textsc{DrawBatch}(\mathcal{B})$} \Comment{urgent first}
  \State $(b, k) \gets \textsc{BestBudget}(C)$ \Comment{$k$: in-band requests}
  \If{$k = |Q|$} \Return $(C, b, L)$ \EndIf
\EndFor
\State $(C, b, k) \gets$ the candidate with the largest $k$

\If{$\exists\, \beta \in \mathcal{B}: \textsc{Context}(\beta) > M$} \Comment{transient burst}
  \State $(C', \ell) \gets \textsc{Widen}(C)$
  \State $(b', k') \gets \textsc{BestBudget}(C')$
  \If{$k' > k$} \Return $(C', b', \ell)$ \EndIf
\EndIf
\State \Return $(C, b, L)$
\end{algorithmic}
\end{algorithm}
 

\PP{Batching the requests} 
Serving requests in arrival order
respects their deadlines when every request carries the same SLO, since
the earliest arrival is the most urgent, but it is not enough. A batch's
deadline is its most urgent member's, while its execution time is set by
all of its members, and a batch formed in arrival order that pairs an
urgent short request with a long one bears the long request's
computation under the short request's deadline. \ours-Scheduler instead
runs requests of similar length and latency slack together: short
requests share a small budget that suffices for them, and the long
request runs alone within its own band.


Since the number of possible compositions grows exponentially with the
queue depth, searching them exhaustively is expensive. Requests of
similar length and deadline, however, are interchangeable: swapping one
for another changes neither the batch's execution time nor its deadline. \ours-Scheduler therefore groups the queue into
buckets by deadline and context length and searches over compositions
that specify only the number of requests drawn from each bucket, most
urgent first within a bucket.
Requests that cannot meet their SLO even at the floor are held in
a bucket of their own. \ours serves them at full reuse, and raises
their coverage as far as the transfer leaves room for free.
 
\PP{Raising throughput transiently} If the serving system is persistently overloaded, \ours may find no schedule that accommodates every request, and reordering alone cannot fix that: it only changes the victim. The request it pulls forward accommodates; the one it displaces no longer does. A transient burst, where requests arrive together and the queue clears once they finish, is different. The limit on those requests is not the order but how many run at once.
 
The KV cache caps how many run at once, and \ours-Scheduler can raise
that cap by holding less of each request resident: under \emph{partial
KV allocation}, a request keeps only the layers about to run resident,
and the rest is loaded as the batch advances. It reads the condition
from the buckets. Requests that arrive in a burst share their remaining
slack and fall into one deadline bucket, and reordering cannot separate
them. When the total context of requests in the same bucket exceeds the
KV capacity, \ours-Scheduler searches again with the effective capacity
that partial KV allocation raises, and takes the composition it finds
when it places more requests.
The decision is made per batch: once the bucket fits again, the next
batch runs fully resident, and a wider batch meanwhile drains the queue
in fewer rounds.
 
\PP{Scheduling overhead} The search costs a few milliseconds per batch
boundary, negligible against the TTFT it schedules. \ours-Scheduler also
caps it at 256 candidate compositions, and since the most urgent buckets
and the floor are evaluated first, the candidates within the cap are the
ones the objective is most sensitive to. In practice, the cap is rarely
reached.

 \begin{figure*}[t]
    \centering
   \includegraphics{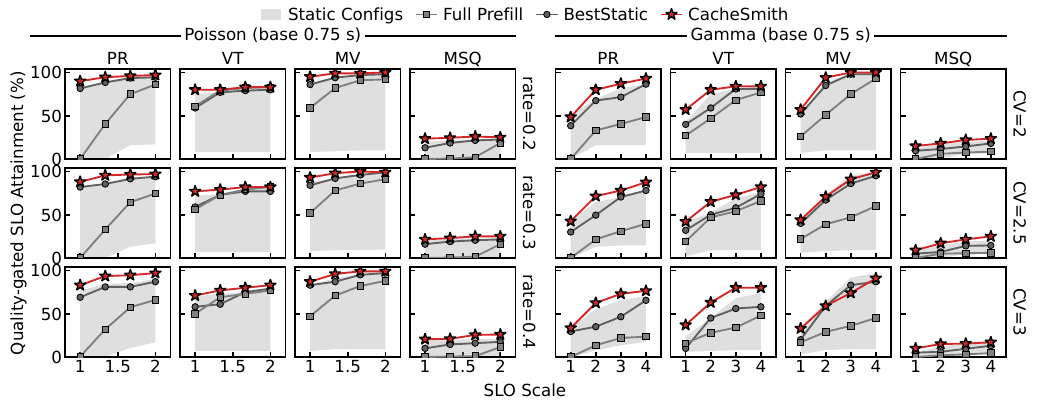}
    \caption{Quality-gated SLO attainment as the SLO target scales, for each task
under Poisson and Gamma arrivals.}
    \label{fig:e2e:grid}
\end{figure*}

\subsection{Serving Runtime}
\label{sec:design-kv}
 
Once the full configuration is decided, it is forwarded to the serving runtime that executes the configuration layer by layer and loads cached KV asynchronously before the layer that reads it. 
 
\PP{Transfer planning} Unlike the layer-by-layer transfer pipelining
that prior work commits to, the KV loader generates a transfer plan
from the $(S, w, p, r)$ configuration once \ours-Planner returns it: a
\warmup layer requires no transfer, a \probe layer its layer's cached
KV, and a \patch layer only the tokens it reuses. The loader then issues
the transfers in order of their deadlines and keeps the link busy as
long as the batch's KV allocation has room.

 
\PP{Partial KV allocation} In case of a transient burst, \ours-Scheduler
may allocate the KV cache partially, for a portion of a request's
layers, in order to admit more requests and drain the waiting queue
faster. The impact on the prefill pass is limited, since a pass never
requires the full cache at once: it writes a layer's KV once and never
revisits it. The KV cache then works as a sliding window over the
layers, and a later layer reuses a slot once the KV that occupied it
has left. A window of any size stalls in two cases. The first is when
every slot is occupied because the transfer off the GPU lags behind.
The second is a \probe under query-context interaction, which rewinds
execution to the last \warmup layer and would need KV already evicted
if the rewind were longer than the window; \ours therefore floors the
window at the deepest \probe it profiles.
The first case is rare in modern serving. Under prefill--decode
disaggregation the KV is sent to the decode instance after prefill in
any case, and partial allocation merely sends it early; the only added
overhead arises when the decode instance, itself busy, rejects it.
Without disaggregation, modern systems still keep a distributed shared
KV store~\cite{mooncake}, and the offload adds no storage, only
transfer. Only in the traditional setup, with prefill and decode
colocated and no shared KV store, does partial allocation add storage
demand, and its offload may contend for the bandwidth over which the
reused KV is loaded; \ours therefore uses partial allocation there only
when serving is not transfer-bound, so that the offload stays off the
critical path.

\begin{figure*}[t]
    \centering
\includegraphics{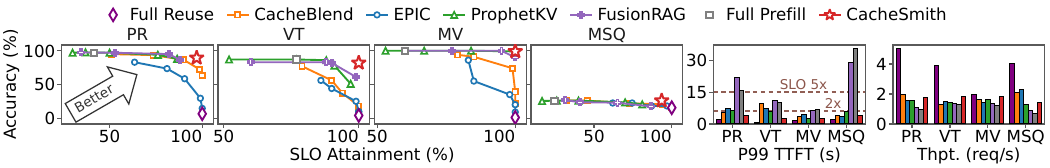}
    \caption{Accuracy against SLO attainment (a--d), and P99 TTFT and throughput
(e, f), at Gamma CV 2.5 and a 3-sec SLO.}
    \label{fig:e2e:frontier}
\end{figure*}

\section{Evaluation}
\label{sec:eval}

We implement \ours on vLLM 0.10.2~\cite{vLLM} with three additional features.
(1)~The \textbf{phase-aware KV transfer engine} loads the cached KV of each request's retrieved
chunks into the cache blocks that the scheduler assigns it. It issues transfers
that share a deadline concurrently over the device's four I/O streams, which lets
one deadline group use its full bandwidth. (2)~\textbf{Custom Triton kernels}
materialize the attention weights that a query-context \probe reads and that
vLLM's default FlashAttention path never writes out. Since the published
implementations' kernels are unavailable, \ours and the query-context baselines
share ours, which is untuned, and latencies under that signal are pessimistic
relative to the published numbers. (3)~\textbf{Partial KV allocation} lets a
later layer reuse the cache blocks an earlier one held. Before it runs, it asks
the transfer engine whether the KV it reads has arrived and whether those blocks
have been released.

\subsection{Experimental Setup}
\label{sec:eval:setup}
 
\PP{Hardware and models} We evaluate \ours on NVIDIA RTX PRO 6000 (96 GB) GPUs, an AMD EPYC 9554 CPU and a Samsung 9100 PRO NVMe SSD over PCIe Gen5$\times$4. We serve Llama-3.1-8B-Instruct on one GPU and Qwen3-32B on two.

\PP{Workloads} We draw requests from four tasks: PassageRetrieval (PR) and MuSiQue (MSQ) from LongBench~\cite{longbench}, and VariableTracking (VT) and Multi-valueRetrieval (MV) generated following RULER~\cite{ruler}. 
PR and MSQ contribute 200 requests each, VT and MV 100. MSQ's passages
are also the longest and the most widely spread, reaching 15K tokens. A
mixed workload (MIX) draws 100 requests from each task, sampling PR and MSQ at
random. MIX approximates a deployment that serves several tasks at once, over a
wider spread of context lengths than any one task presents.
Each passage is one chunk, prefilled in isolation before serving begins and persisted to the SSD. The stores range from 39\,GB for VT under Llama-3.1-8B to 707\,GB for MSQ under Qwen3-32B. The bandwidth study serves the store from host memory at 10\,GB/s and from the SSD at 6 and 3\,GB/s, the two slower rungs obtained by capping the link. 
 
\PP{Traces} We shuffle each workload and replay it as an arrival process. Poisson
arrivals run at 0.2 to 0.4 requests per second. Gamma arrivals run at 0.3 requests per second, with \emph{coefficient of
variation} (CV) 2, 2.5 and 3.

\PP{Metrics} Our primary metric is quality-gated SLO attainment
(QSA; \autoref{eq:qsa}), the fraction of requests answered both within their TTFT SLO
and above their quality gate. We score PR and MSQ by exact match, which makes their gate binary, and VT and MV
by F1, where a request passes above 0.5. We set the base SLO to 0.75\,sec, the mean TTFT of
Full Prefill measured on a single request across the four tasks, and
scale it over 1--2$\times$ under Poisson arrivals and 1--4$\times$ under Gamma arrivals.

\PP{Baselines} We compare against four selective recomputation methods, each
run over a range of coverages---CacheBlend~\cite{cacheblend}, EPIC~\cite{epic},
ProphetKV~\cite{prophetkv} and FusionRAG~\cite{fusionrag}---together with Full
Reuse and Full Prefill as the extremes of the design space. 
Every method runs on the same serving stack, and the transfer engine's
optimizations apply to all of them uniformly. We additionally report an oracle we call \emph{BestStatic}:
for each SLO and arrival pattern, the single configuration that
maximizes QSA across the four tasks, chosen from the
measured results. The oracle is thus granted the SLO, the arrival pattern, and
the outcome itself; only the task is withheld, since it is not known online.
\begin{figure}[t]
    \centering
    \includegraphics[width=\linewidth]{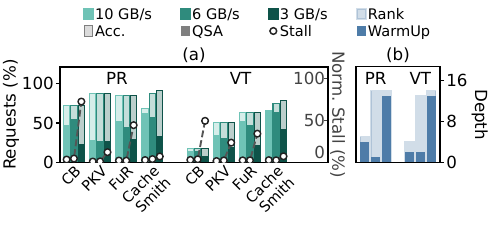}
    \caption{(a) QSA, accuracy and stall with decreasing link speed and (b) the ranking depth \ours-Planner selects. }
    \label{fig:e2e:tier}
\end{figure}

\subsection{End-to-End Performance}
\label{sec:eval:e2e}
 

\PP{Quality-gated SLO attainment} \autoref{fig:e2e:grid} reports QSA over four tasks, six arrival patterns and four SLO targets---96 settings in total. The band spans every static configuration on each task. We observe that \ours exceeds BestStatic in 93 / 96 settings and improves QSA by {2--27.5}\%. 
The penalty of a misconfiguration is significant: BestStatic achieves up to 88\% higher QSA than the worst configuration. 
BestStatic does not reach the band's top in most panels: the best configuration is also task-dependent, and the task is not known in advance. 
Notably, even BestStatic achieves worse QSA than Full Prefill in 8 / 96 settings, where queueing is mild and the SLO is loose enough that Full Prefill meets its target; \ours is above Full Prefill in all evaluations.

 
\PP{Source of the gain} \ours's gain comes from co-optimizing accuracy and the
SLO rather than from leading on either alone. \autoref{fig:e2e:frontier}(a--d)
plots accuracy against SLO attainment on PR, VT, MV, and MSQ. \ours achieves neither
the highest accuracy nor the highest attainment, yet reaches 97\% attainment at
89\% accuracy on PR, a combination no prior method reaches at any coverage.
\autoref{fig:e2e:frontier}(e,\,f) report P99 TTFT and throughput at each
method's highest-QSA configuration. \ours has the shortest tail of any method
except Full Reuse, and throughput comparable to the rest. FusionRAG alone shows
a transfer stall: its query-only pilot pass is unable to hide the KV transfer, pushing its P99 to 22\,s on PR.

\PP{Transfer-aware planning} \ours adapts to the transfer bandwidth available to
it, because \ours-Planner prices a plan on transfer as well as computation. 
We host the KV store in host memory and on disk, at measured bandwidths of 10, 6
and 3\,GB/s, and report accuracy alongside QSA (\autoref{fig:e2e:tier}(a)). As
the bandwidth drops, a static configuration cannot cover the KV transfer it must
wait on, and baselines spend 12--72\% of the serving time stalled. The stall eventually translates into QSA degradation. In contrast, \ours spends only 3--7\% of that time stalled, and its accuracy
rises as the link slows, from 69\% to 92\% on PR and 66\% to 78\% on VT.
\ours-Planner achieves this with transfer-aware planning
(\autoref{fig:e2e:tier}(b)). At 6\,GB/s it reaches a rank depth of 14 almost
entirely by probing, the route that costs the least computation. At 3\,GB/s it
reaches the same depth with 13 \warmup layers and a single \probe, since a
\warmup reads nothing from the store and the recomputation it performs while
cached KV is in flight costs no latency.

\subsection{Ablation}
\label{sec:eval:ablation}
 \begin{figure}[t]
    \centering
    \includegraphics{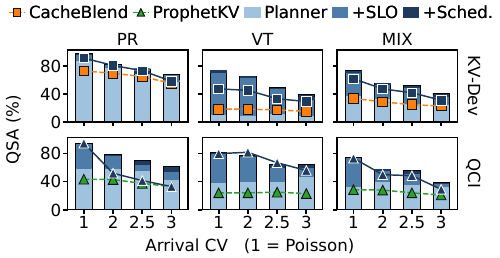}
    \caption{Contribution of each component and baseline enhanced by \ours-Scheduler. }
    \label{fig:val:decomp}
\end{figure}

 
\PP{Component contribution} We compare \ours within each selection-signal family,
against CacheBlend for KV deviation and ProphetKV for query-context interaction
(\autoref{fig:val:decomp}). \ours-Scheduler generalizes beyond \ours: scheduling
CacheBlend under it raises QSA by 10\% on PR and 20\% on VT, and ProphetKV by
29\%, with only their coverage varying to meet the budget it assigns.
\ours-Planner alone runs at a fixed budget matched to the baseline's and exceeds
it in most cells. \emph{+SLO} replaces
that fixed budget with one derived from the deadlines of the whole queue, and at
CV 1 it accounts for nearly the entire improvement. \emph{+Sched.} adds partial KV allocation and out-of-order scheduling, worth at
most 2\% at CV 1 but 7.5\% at CV 3, where bursts deepen the queue and the SLO-aware budgeting alone no longer suffices.

\begin{figure}[t]
    \centering
    \includegraphics{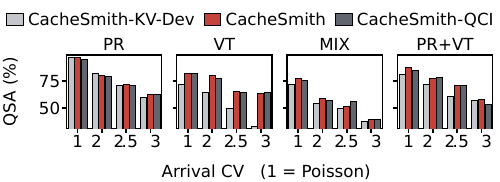}
    \caption{\ours compared with \ours with fixed signals.}
    \label{fig:val:signal}
    \vspace{-1em}
\end{figure}

\PP{Selection signal} We compare \ours against two variants that fix the
selection signal, on PR and VT alone, on MIX as a realistic multi-task
deployment, and on PR+VT, which shuffles the two opposing tasks into an extreme
case (\autoref{fig:val:signal}). On PR and VT nearly every request favors one
signal, and \ours matches the suited variant rather than beating it: on VT, KV
deviation alone reaches 33\% QSA at CV 3 against 63\% for \ours, while on PR the
three stay within 3\%. \ours exceeds both fixed variants where a workload holds
both preferences, on MIX and on PR+VT. That gain shrinks as arrivals become
burstier, since each batch adopts the majority signal among its requests and
larger batches co-batch opposing preferences more often, leaving the minority to
run under the wrong signal.
 
\begin{figure}[t]
    \centering
    \includegraphics[width=\linewidth]{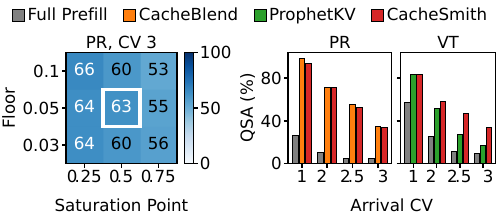}
\caption{(a) QSA on PR at CV 3 across budget floor and saturation point (box: default).
(b) QSA on Qwen3-32B (TP=2) across burstiness against Full Prefill and BestStatic configuration (CacheBlend 20\% on PR, ProphetKV 20\% on VT).}
    \label{fig:val:band}
    \vspace{-1.5em}
\end{figure}

\PP{Budget band} \ours-Planner delivers \ours-Scheduler a band between a floor
and a saturation point, 5\% and 50\% coverage by default. Raising the saturation point to 75\% costs
4\% to 8\% QSA on PR at CV 3, since a batch then spends budget that no longer
helps it while every request behind it waits (\autoref{fig:val:band}(a)). The
floor moves QSA by at most 3\% either way: a higher floor forces a minimum some
requests cannot afford, sending them to Full Reuse, and a lower one admits
budgets too small to clear the quality gate.

\PP{A larger model} 
We evaluate it on Qwen3-32B under two-way tensor parallelism (TP=2), and showcase its potential to generalize to other models without extensive reprofiling the rank-accuracy trade-off. We recalibrated the rank-accuracy distribution profiled on LLaMA-3.1-8B curve with only 20\% from the full set and interpolated the rest, and tuned only the budget band to match. 
Deploying \ours with tensor parallelism requires only adding all-reduce communication for \probe layers, at negligible cost. \ours exceeds the BestStatic method on VT, by 20\% at CV 2.5 and 17\% at
CV 3, and is on par on PR, within 4\% at every arrival pattern
(\autoref{fig:val:band}(b)). Full Prefill trails both on either task as soon as
arrivals become bursty, reaching 4\% on PR at CV 3.


\section{Related Work}
\label{sec:related}

\PP{RAG context caching and KV reuse}
Prior systems cut RAG prefill cost by
caching precomputed KV across memory hierarchy and
sharing it across queries or instances~\cite{ragcache, cachegen, lmcache, mooncake, RAG-DCache}.
They decide where reusable KV lives and how it moves, not the impact of cross-chunk attention; \ours sits above them and makes that call
under a budget. Others avoid the loss by fine-tuning the model to attend within
chunks or reconnect them~\cite{blockattention, turborag, kvlink}, which removes
online recomputation but requires modifying the served model.

\PP{SLO-aware LLM serving}
Serving systems schedule requests to meet latency targets:
Orca~\cite{orca} batches continuously, Sarathi-Serve~\cite{sarathi-serve} interleaves
chunked prefill with decode, QLM~\cite{qlm} manages the queue against head-of-line
blocking, and Andes~\cite{andes} preempts to maximize token-delivery QoE.
Disaggregated designs~\cite{distserve, splitwise} give each phase its own SLO, a
setting a prefill-side system fits naturally. Others offload KV and overlap its
transfer with compute~\cite{infinigen, flexgen}; OrbitFlow~\cite{orbitflow}
reconfigures per-request KV placement during decode, complementary to our
placement of recomputation during prefill. All schedule a fixed amount of work
per request. \ours makes that amount a decision variable, setting a batch's
budget together with its composition.

\PP{Quality-aware serving}
Systems trading accuracy for latency select among models or configurations:
Proteus~\cite{proteus} serves cheaper variants under load, Tabi~\cite{tabi}
cascades queries through multiple levels, MFS~\cite{mfs} serves a model family
jointly, and Metis~\cite{metis} adapts RAG pipeline configurations to quality
targets. \ours shares the objective but varies how much reused KV is
recomputed within one model, so no model switch sits on the critical path.

\PP{RAG serving systems}
Beyond the prefill stage, RAGO~\cite{rago} searches over task placement,
resource allocation, and batching across retrieval and generation,
PipeRAG~\cite{piperag} overlaps iterative retrieval with generation, and others
partition resources between the vector index and the LLM~\cite{hedrarag,
vectorliterag, chameleon}. These systems govern how pipeline stages share a machine;
\ours optimizes the prefill stage within them.

\section{Conclusion}
\label{sec:conclusion}

Context caching accelerates RAG prefill, but existing selective recomputation
schemes enforce rigid configurations that fail under dynamic latency budgets
and heterogeneous requests. We present \ours, a serving system that
decomposes selective recomputation into a modular workflow. By pairing an
SLO-aware scheduler with a budget-aware planner, \ours is the first system
that coordinates batching and dynamic execution planning across workflow
stages to maximize quality-gated SLO attainment.

\bibliographystyle{ACM-Reference-Format}
\bibliography{refs}

\end{document}